\documentclass[msom,nonblindrev]{informs4_copy} % current default for manuscript submission

\OneAndAHalfSpacedXI % current default line spacing
\usepackage{changes}
\usepackage{placeins}  % Add this to your preamble
\usepackage{tikz}
\usetikzlibrary{calc}
\usepackage{pgfplots}
\pgfplotsset{compat=1.15}
\usepackage{caption}
\usepackage{comment}
\usepackage{enumitem}
\usepackage{xcolor,soul}
\sethlcolor{lightgray}
\usepackage{multirow}

\definecolor{lightgray}{gray}{0.9}
\newcounter{marginparcounter}

\usepackage{subcaption}
\usepackage{booktabs}
\usepackage{arydshln}
\usepackage{natbib}
 \bibpunct[, ]{(}{)}{,}{a}{}{,}%
 \def\bibfont{\small}%
 \def\BIBand{and}%

\newcolumntype{R}[1]{>{\raggedleft\let\newline\\\arraybackslash\hspace{0pt}}m{#1}}
\newcolumntype{C}[1]{>{\centering\let\newline\\\arraybackslash\hspace{0pt}}m{#1}}
\newcolumntype{L}[1]{>{\raggedright\let\newline\\\arraybackslash\hspace{0pt}}m{#1}}

\TheoremsNumberedThrough     % Preferred (Theorem 1, Lemma 1, Theorem 2)
\ECRepeatTheorems

\EquationsNumberedThrough    % Default: (1), (2), ...
\MANUSCRIPTNO{MS-000X-XXXX.XX}

\begin{document}
%%%%%%%%%%%%%%%%

%\setcounter{page}{1}
%\DOI{10.1287/xxxx.0000.0000}%

\RUNAUTHOR{Kagan, Jost, Schiffels and Lieberum}

\RUNTITLE{Does Iteration Drive Innovation?}

% \TITLE{Bundling Information Goods of Decreasing Value}
\TITLE{On Repeat: Does Iteration Drive Innovation?} %[Note: Vielleicht finden wir zusammen noch was besseres. Ich finde The Strengths and Weaknesses passt klingt sehr konkret und allgemein. auf generelle stärken und schwächen gehen wir im Paper aber nur wenig ein. primär zeigen wir dass agile zwar gut aber eben auch schlecht sein kann.]

\ARTICLEAUTHORS{%
\AUTHOR{Evgeny Kagan}
\AFF{Johns Hopkins Carey Business School, Johns Hopkins University, \EMAIL{ekagan@jhu.edu}}
\AUTHOR{Christian Jost}
\AFF{University of Augsburg,  \EMAIL{christian1.jost@uni-a.de}}
\AUTHOR{Sebastian Schiffels}
\AFF{University of Augsburg, \EMAIL{sebastian.schiffels@uni-a.de}}
\AUTHOR{Tobias Lieberum}
\AFF{Unaffiliated, \EMAIL{tobias.lieberum@tum.de}} 
}

\ABSTRACT{\textbf{\textit{Problem definition:}} Motivated by the widespread adoption of iterative project management techniques, we study the effects of workflow -- iterative or sequential -- on innovative behavior and performance.  \textbf{\textit{Methodology/Results:}} We conduct a series of laboratory experiments. Our first experiment shows that, in an open-ended creative challenge, iterative task completion leads to better outcomes than sequential task completion. In the second experiment we show that the advantage of iterative workflow further extends to innovation settings that do not involve idea generation. A key mechanism driving the advantage of iterative work is that it leads to frequent task switching,  prompting workers to perform a broader search for the best available solution. In the third experiment we delve deeper into the search process and show that sequential work indeed leads to more myopic idea refinement behaviors, often ending in a (suboptimal) local maximum. \textbf{\textit{Managerial implications:}} Our results suggest that iterative workflow improves performance across multiple, structurally distinct innovation settings. We also identify three boundary conditions. First, iterative workflow helps achieve quick gains, but its performance advantage narrows over time. Therefore, workflow effects are stronger when balanced performance across project components is required, but weaker when excellence in one component can offset poor performance in others. Second, workflow has minimal effect on performance in tasks that do not require the worker to perform broad exploration. Third, workflow effects are minimal when workers complete the easier component first.}
  
%\KEYWORDS{project management; innovation; behavioral operations; real-effort experiments} 

\maketitle

\vspace{-0.2cm}
\section{Introduction}
Effective time management is central to most innovation activities. The way workers navigate tasks -- sequentially, focusing on one task at a time or \textit{iteratively}, working on multiple tasks concurrently and completing them in increments rather than whole -- can greatly affect efficiency, the quality of output, and the potential to generate and develop ideas. In this paper we report the results of a series of experiments examining the effects of sequential and iterative workflow on innovative behavior and performance.

Our research is motivated by broader trends in knowledge work and innovation management that emphasize autonomy, flexibility, and iterative progression of tasks. A prominent example is the widespread adoption of Agile project management techniques \citep{RIGBY.2016,kettunen2020,allon2021,lieberum2022,ghosh2023}. Our interactions with Allianz SE, a large multinational insurance company, suggest that Agile methodologies have significant consequences for individual workflow.\footnote{The insights presented here are gathered from our interactions with the Chapter Lead of IT Information Security within Allianz’s Global IT Security division.} Allianz’s shift to Agile introduced two-week sprints, each concluding with a prototype demo and planning for subsequent sprints. Within each sprint, workers experience increased task fragmentation and more frequent iterations, with features being developed, tested, and launched in accelerated, overlapping cycles. As a result, workers face increased task switching and are encouraged to exercise greater autonomy in prioritizing tasks rather than completing them in a fixed, predetermined order. For example, when developing software applications, a developer may now frequently switch between front- and back-end components within the same sprint -- and often within the same day -- ensuring parallel progress on both components. This contrasts with the previously used Waterfall approach, where back-end development would typically precede front-end design.

While Agile methodologies can promote frequent, autonomous task switching and parallel progress on multiple tasks, popular management concepts such as "Deep Work" \citep{newport2016} and "Flow" \citep{csikszentmihalyi2013} suggest the opposite: minimal task switching and prolonged sustained focus. These frameworks suggest that innovative work requires extended periods of uninterrupted concentration. A common implementation of these concepts involves scheduling dedicated blocks of focused work on a single task. Creative teams at advertising agencies like Ogilvy and tech companies such as Google implement this practice through structured work sessions to maintain high levels of concentration \citep{eyal2014, cirillo2018}. These management concepts are also supported by some psychological and neuroscience research that suggests that task-switching can lead to slower reaction times and higher error rates \citep{monsell2003,kiesel2010}.

The above examples suggest a tension between iterative approaches, such as Agile, which emphasize more flexible, worker-determined workflow with frequent, unplanned reordering of tasks and priorities, and sequential approaches that emphasize deliberate reduction of task switching. We seek to inform this debate by reporting the results of a series of experiments that examine how innovation performance is affected by two prototypical workflows underlying many project management frameworks: (1) a \textit{sequential workflow} where each task is completed within a single allocated time block, and (2) an \textit{iterative workflow} where each task is completed in multiple worker-determined increments. We focus specifically on innovation contexts characterized by the following four features:

\begin{enumerate}
\item \textbf{Each worker handles multiple tasks.} We focus on settings where the workflow affects not only how teams divide and coordinate tasks between workers, but also how individual workers allocate their time between tasks. That is, workers are cross-trained to work on different features or components of a product rather than being assigned to a single feature or component. For example, at Allianz, software developers often work on both front-end and back-end system components during the same sprint. 
\item \textbf{Task-switching costs are low.} The tasks assigned to a worker are within the same technological domain. That is, the tasks are similar enough that switching between them does not require significant retooling or incur substantial setup costs. 
\item \textbf{Rework can be costly.} As development progresses, the costs associated with rework increase, particularly for tasks or components that have been completed early on in development. Such cost increases are well documented in the literature \citep{terwiesch1999,loch1999}. Our partners at Allianz have frequently observed that a major drawback of the Waterfall methodology is its tendency to require  such costly rework at the end of the project. 
\item \textbf{Performance is limited by the lowest-performing component.} The overall outcome relies on each component meeting certain performance levels.  In our discussions with Allianz, such projects were the norm rather than the exception, and were referred to as ``weakest-link'' projects -- meaning that excellent performance in one component cannot compensate for poor performance in another (In follow-up experiments in \textsection 7 we will examine an alternative setting where this assumption is relaxed). 
\end{enumerate}

To study the effects of iterative versus sequential workflow on innovative behavior and performance, we conduct a series of laboratory experiments in which we carefully vary the workflow (iterative vs. sequential) in several structurally distinct innovation settings. Our experimental design draws on the experimental psychology tradition \citep{guilford1950, torrance1966, simonton2000, sawyer2011}, which focuses largely on idea generation and brainstorming stages of innovation activities. Additionally, we leverage the innovation and search literature in economics and management \citep{levinthal1981, levinthal1997, mihm2003, ederer2013, billinger2014, sommer2020}, which focuses on the more downstream stages of innovation. Our experiments examine behaviors related to both the generative stages where ideas are created, as well as behaviors related to the later stages of innovation where the best ideas are selected and combined into a single integrated whole.

Our first experiment is an open-ended creative challenge. The experimental task is a variation on the popular game ``Scrabble'' -- participants are given letters of the alphabet and must build connected verbal structures under time and material constraints. The Scrabble task consists of two components: one component is a Scrabble board where participants can only use verbs, while the second component is a board where they can only use nouns. In the sequential workflow treatment participants first complete one task (e.g., nouns), and then move on to the second one (e.g., verbs). In contrast, with an iterative workflow participants split their time into shorter increments as they work on both tasks (nouns and verbs), repeatedly switching between them as they see fit. 

Our main experimental result is that iterative workflow outperforms sequential, with performance improvements ranging between 15 and 28 percentage points. In additional treatments we rule out the explanation that iterative workflow allows workers to spend more time on the more difficult or value-adding task. Indeed, participants in all treatments spend approximately the same amount of time on each task. That is, improved productivity is caused not by better time allocation, but by frequent task switching and refocusing one's attention on a new problem. Examining participant productivity, we find that the low performance of the sequential treatment is driven mainly by sharply diminishing marginal gains as workers approach the end of the time period allocated to each task. This is especially true during the initial work period, suggesting that participants run out of ideas faster with sequential workflow, especially when the task is new to them and they are required to generate new ideas in an unfamiliar environment. In contrast, iterative workflow appears to stimulate creative production at a more steady pace.

The Scrabble task involves both creative idea generation (forming words), and a more integrative, combinatorial element (connecting new words with existing ones). 
To examine whether the advantage of iterative work extends to settings that lack the idea generation component, we conduct a second experiment in which participants are supplied with a set of ideas rather than having to generate new ideas. Similar to the first experimental task, this task is based on Scrabble. However, different from the classic Scrabble task, participants now receive a list of \textit{pre-formed} words, and the task is to use as many of these words as possible to build a single connected structure. We show that in this setting iterative workflow continues to outperform sequential, suggesting that the performance gap is not driven solely by differences in idea generation and creative production, but also extends to more downstream innovation activities.

A key performance indicator in the second experiment is the number of restarts of the search for new solutions (participants removing words from the board and starting from scratch). Indeed, such restarts occur more frequently with iterative workflow, suggesting that iterative workflow stimulates more exploratory behaviors, while sequential workflow leads to more narrow and myopic strategies. To better understand this mechanism, we introduce a third experiment in which we use a version of the multidimensional search task \citep[``Lemonade Stand Task'',][]{ederer2013,sommer2020}, and zoom into explore-exploit behaviors as a potential driver of the advantages of iterative work.  Comparing search behavior and performance across different parametrizations of the search landscape, we find that iterative workflow is beneficial on more ``rugged'' (or interdependent) landscapes where it leads to a broader exploration of the solution landscape. In contrast, sequential workflow performs just as well as iterative on smoother landscapes where the ``greedy'', myopic hill-climbing approach is a viable strategy. 

In addition to search landscape ``ruggedness'', we are able to identify two further boundary conditions on the benefits of iterative workflow. First, iterative workflow helps achieve quick gains, but its performance advantage narrows (or even reverses) over time. This is driven by workers in sequential workflow regimes significantly improving their performance in the second period relative to the first. Therefore, workflow effects are minimal when good performance in one component can offset poor performance in others. Second, we show that the effects of workflow are minimal when workers are allowed to complete the easier component first.  

\section{Literature and Contributions}
Innovation and creativity research spans various fields, including psychology, economics, strategy, and operations management \citep{krishnan2001,sorensen2010,sawyer2011,kavadias2020}. However, there remains a significant gap in our understanding of how workflow affects innovation behavior and performance. No study, to our knowledge, examines the micro-level dynamics of task management in innovation-related tasks. We next discuss four streams of literature that inform our experiment design, and that our study contributes to: the task selection literature, the task-switching literature, the project management literature, as well as the broader innovation research that uses real-effort tasks.  

\subsection{Task Ordering and Selection in Operations Management}
The study of people-centric operations, i.e., of how human factors influence operational processes, has gained attention in recent years \citep{roels2021}. A notable stream within this literature is research on task ordering and selection, conducted mainly in medical settings. \cite{ibanez2018} study physicians' task prioritization and observe a tendency to choose the shortest or easiest tasks first. This finding is further validated by \cite{kc2020}, who show that such behaviors negatively affect performance in both lab and field. If these patterns extend to our innovation setting, we should expect  suboptimal performance with the more flexible, iterative workflow, as workers may allocate more of their time to less complex tasks, rather than to those contributing most value. To ensure that workers do not overspend time on easier or more enjoyable task  components, one of our iterative workflow treatments restricts the time spent on each task to be the same.  \cite{kc2011} use hospital data and show positive effects of focus on performance. \cite{narayanan2009} find similar effects in a software support setting. Similarly, \cite{staats2012} find that task variety can have negative effects in the short term, which would also speak for sequential workflow  in our setting. \cite{kc2014} finds that some task switching can enhance productivity and quality of care, but also cautions against its overuse. None of these studies examine innovation-related activities. Finally, \cite{siemsen2008} and \cite{katok2011} examine task selection in principal-agent settings where workers use the difficulty level of tasks to signal ability  and garner greater rewards. Our experiments abstract away from any interactions between workers and managers, and instead use individual tasks with objectively measurable performance. 

\subsection{Task-switching in Psychology}
The task-switching paradigm, well-established in psychology and cognitive sciences, examines how individuals respond to different task sequences. Subjects are typically presented with a sequence of short, unrelated tasks \citep{vandierendonck2010}. Switching refers to whether the next task is of the same type (repeat) or a different type (switch), with sequences like A A B B (repeat, switch, repeat) or A B A B (switch, switch, switch). A classic finding in these studies is that more frequent switching negatively affects performance \citep{kiesel2010}. A commonly proposed mechanism is that switching makes it difficult to fully transition one's attention, which leads to reduced performance in both the interrupted and interrupting tasks \citep{leroy2009,leroy2020}. 

Our approach differs from the classic task-switching paradigm in several ways. First, the literature typically focuses on moving between very different tasks that engage different cognitive resources \citep{kiesel2010}. In contrast, in project management settings, tasks are often related, as workers complete tasks within the same project or between closely connected projects. In such contexts, and in our experiment, workers engage in \textit{component switching} (or \textit{feature switching}) rather than switching between entirely unrelated tasks. Second, most psychological experiments involve a sequence of short, pure effort-based tasks, such as categorizing numbers as odd or even or letters as consonants or vowels. In contrast, we focus on more complex, innovation-related tasks. Third, much of the literature does not separately identify the effects of autonomy from the effects of task switching, both of which are integral to many iterative task management techniques. Most studies impose a fixed task sequence (e.g., A A B B) with a fixed amount of time for each task.\footnote{There are several exceptions to this, typically referred to as \textit{voluntary} or \textit{discretionary} task switching; see \cite{kiesel2010}.} In contrast, we allow participants to choose \textit{when} to switch between tasks, which is more reflective of real-world project management practices. Finally, instead of presenting participants with separate, independent performance rewards for each sub-task, our tasks are integrated and require participants to work towards a common, project-based performance reward.\footnote{In \textsection 7, we examine an alternative incentive system in which performance is evaluated based on the sum of component scores. This incentive system is more in line with the task-switching literature in psychology.}

The closest study to ours is \cite{lu2017}, who examine discretionary task switching using classic psychological tasks such as the Alternative Uses Task \citep[AUT;][]{guilford1950} and the Remote Associates Test \citep[RAT;][]{mednick1968}. Unlike our study, these tasks are short and involve either idea generation alone (AUT) or arriving at a single correct insight (RAT), which are quite different from the more complex, integrated innovation tasks we use.

\subsection{Project Management and Time Allocation}
The closest related project management studies are \cite{kagan2018} and \cite{lieberum2022}. \cite{kagan2018} find that workers who decide for themselves how to spend time between creative ideation and execution perform worse than workers with exogenously imposed schedules -- an effect driven mainly by delays in worker-determined schedules.  \cite{bendoly2014} find that managerial progress checks are key to effective task switching.  \cite{lieberum2022} show that time-boxing of work, i.e., imposing fixed time intervals for tasks, can improve performance. They use a slider task, which measures pure effort (as opposed to innovation performance). While these studies examine work arrangements that give workers more/less process control and autonomy, none of them compare iterative vs. sequential workflow, or explore multiple, structurally distinct, innovation settings.

\subsection{Broader Literature on Innovation and Creativity}
Two of our experimental tasks build on the experimental psychology tradition of using verbal tasks to study creative behaviors \citep{sawyer2011}. A common approach is to use verbal puzzles or riddles  \citep{kachelmeier2008,erat2016} or deciphering anagrams \citep{mendelsohn1964, ansburg2003}. Many of these tasks emphasize the creative ideation component of the innovation process, where the objective is to produce as many ideas as possible. One of our experimental activities is based on the popular game ``Scrabble''. Similar to the literature, this game also requires participants to generate many ideas; however, in addition to idea generation, it also reproduces the more downstream stages of innovation, such as the need to select, combine and implement the best ideas into a single integrated product. 

The idea generation literature also addresses techniques for bolstering creativity, with particular attention to ``incubation periods'' -- strategic breaks in creative work \citep{helie2010,ritter2014}. Within our setting, an iterative workflow effectively functions as such an incubation period. By periodically shifting attention away from one task and focusing on a different one, participants temporarily step back from their current problem. In this way, our findings contribute to the incubation literature by demonstrating that managing creativity through structured iterations can improve creative production.

Our third experimental task leverages the approach (more common among economists and business disciplines) of representing innovation as a multidimensional search process \citep{levinthal1981, levinthal1997,mihm2003,sommer2004}. In this setting, idea generation is muted; instead, each potential solution is a vector of product attributes, and the worker's goal is to identify the best combination among a very large number of possibilities, typically under time constraints. To achieve good performance the worker needs to develop an understanding of the mapping between combinations of product attributes and the resulting performance.  While the theoretical  literature on complex solution landscapes  is exhaustive \citep[in particular for NK models; see][for a recent review]{baumann2019}, the number of experiments examining human search strategies on a landscape is relatively small. See \cite{ederer2013} and \cite{sommer2020} for recent examples. A key advantage of the Lemonade Stand task is that it allows the experimenter to manipulate landscape ``ruggedness''  -- a key moderator of the effects of workflow on performance in our setting.

\subsection{Contributions}
Our study is the first systematic effort that we are aware of, to explore how individual workflow influences innovation behaviors and outcomes across multiple, structurally distinct innovation settings. We contribute to the task selection and project management literature by providing a novel test of key project management techniques within an innovation setting. In addition, our work advances two strands of the broader problem-solving literature: one related to task-switching, by identifying conditions under which task-switching can improve performance in innovation settings, and a second one related to the role of interruptions or ``incubation'' periods in creative processes, by showing how certain project management techniques naturally create incubation periods that promote innovative behaviors and improve performance.

\section{Experiment overview}
To examine the effects of workflow on innovative behavior and performance we conducted a series of laboratory experiments. The experiments were organized into an 8 (\textit{treatments}, between-subject) $\times$ 2 (\textit{tasks}, within-subject) design. Table \ref{tab:design} summarizes the treatments and tasks. %While the details of the treatments and experimental activities will be discussed in \textsection 4-6, here we offer a brief overview and describe the general setup and protocols. 

\begin{table}[tb]
\TABLE
{Experiment Overview\label{tab:design}}
{\small
  \centering
  \renewcommand{\arraystretch}{1.3}
    \begin{tabular}{L{6cm}|L{6cm}|L{5.2cm}}
    \toprule
     \textbf{Workflow} \newline(Treatments varied between-subject) & \multicolumn{2}{C{11cm}}{\textbf{Questions and tasks} (Each subject completes two tasks: a version of the Scrabble game and a version of the Lemonade Stand game)} \\
\midrule   & \textit{\textbf{\textsection 4: How does workflow affect performance in a setting that has both a creative and a combinatorial element? }} & \textit{\textbf{\textsection 6: Do the treatment effects replicate in a different innovation setting?  }} \\  
          & \textbf{Task: Scrabble } & \textbf{Task: Lem. Stand (rugged landscape)} \\  
          \hdashline
    T1: Sequential & \textit{SEQ} & \textit{SEQ} \\
              \hdashline
    T2:   Iterative, time and process constraints & \textit{ITER EQUAL FREEZE } & \textit{ITER EQUAL  FREEZE} \\
              \hdashline
    T3:   Iterative, process constraints & \textit{ITER FREEZE} & \textit{ITER FREEZE} \\          \hdashline
    T4:   Iterative, no constraints & \textit{ITER} & \textit{ITER} \\
              \hdashline
           &       &  \\
           &  \textit{\textbf{\textsection 5: How does workflow affect performance in a setting that has \underline{only} the combinatorial element?  }} & \textit{\textbf{\textsection 6 (cont'd): Does myopic, ``greedy'' optimizing drive the disadvantage of sequential workflow?}} \\
          & \textbf{Task: Scrabble with pre-formed words} &  \textbf{Task: Lem. Stand (smooth landscape)}\\  
                    \hdashline
   T5:  Sequential & \textit{SEQ} & \textit{SEQ} \\
           \hdashline
   T6:  Iterative, no constraints & \textit{ITER} & \textit{ITER} \\
                                    & \multicolumn{2}{C{11cm}}{ } \\
  & \multicolumn{2}{C{11cm}}{\textit{\textbf{\textsection 7.1: How do incentives affect performance? }}} \\
          & \textbf{Task: Scrabble } & \textbf{Task: Lem. Stand (rugged landscape)} \\  
          \hdashline
   T7:  Sequential & \textit{SEQ ALT} & \textit{SEQ ALT} \\
           \hdashline
   T8:  Iterative, no constraints & \textit{ITER ALT} & \textit{ITER ALT} \\
    \bottomrule
\end{tabular}
}
{\textit{Notes.} \textit{ITER} stands for iterative workflow, \textit{SEQ} stands for sequential workflow. \textit{EQUAL} means that the total amount of time allocated to component~1 must be equal to the time allocated to component~2. \textit{FREEZE} means that the choices made in period 1 cannot be altered in period 2. Ordering of activities (Scrabble $\longrightarrow$ Lemonade Stand or Lemonade Stand $\longrightarrow$ Scrabble) was randomized in the experiment. In \textit{SEQ ALT} and \textit{ITER ALT}, subjects were incentivized based on the sum of component scores, as an alternative to the minimum of component scores used in the remaining treatments.}
\end{table}

\subsection{Treatments and Tasks}
In each treatment subjects completed two tasks, administered in random sequence: a version of the   Scrabble task, and a version of the Lemonade Stand task.  The \textit{Scrabble}  task requires participants to engage in both creative idea generation (forming words), and idea recombination and integration (combining words in a performance-maximizing manner). We examine this task in \textsection 4. The \textit{Scrabble with pre-formed words} task removes the creative idea generation element but retains the recombination element. We examine this task in \textsection 5. The \textit{Lemonade Stand} task is an experimental game designed specifically to study innovative behaviors related to idea recombination  \citep{ederer2013,sommer2020}.  The advantage of this task is that it allows a more tightly controlled test of treatment effects on explore-exploit behaviors. In particular, we examine a more rugged landscape where broader exploration is needed to achieve good results, and a smoother landscape where narrower, myopic search can be sufficient. We examine this task in \textsection 6.

Within each task subjects worked on two components. In particular, the Scrabble task included a Scrabble board for verbs only, and a Scrabble board for nouns only. Analogously, the Lemonade Stand task included two separate search landscapes, one related to market attributes of the lemonade stand, and a second one related to product attributes. Depending on the treatment, subjects were either required to complete the components in a pre-determined sequence (\textit{SEQ}), or were allowed to switch between components, thereby completing them iteratively (\textit{ITER}). In addition, we examined two intermediate regimes, in which subjects worked iteratively, but were not allowed to alter the work performed in the first period (\textit{ITER FREEZE}), as well as a regime in which they worked iteratively, but were required to spend equal amounts of time in each component (\textit{ITER EQUAL FREEZE}). Both of these treatments impose constraints on the iterative workflow, bringing it closer to a sequential workflow and allowing us to identify some of the key mechanisms.  

In treatments T1-T6 (\textsection 4-6), participants were incentivized based on the lower of the two components. This was done to model complementarities across components common in many innovation and product development contexts. In treatments T7-T8 (\textsection 7) we examine an alternative incentive scheme, wherein subjects were paid based on the sum of the two component scores, rather than based on the lower score. 

\subsection{Payments and Protocols}
All experiments were conducted at a large public German university. The experiments were programmed and conducted in German, the first language of most of the participants. Sessions consisted of ten to twelve participants. The sequence of tasks (Scrabble $\longrightarrow$ Lemonade Stand or vice versa) and the first displayed component (Verbs $\longrightarrow$ Nouns or vice versa, and Product component $\longrightarrow$ Market component or vice versa) were randomized to control for order effects and fatigue. Participants were paid a fixed show-up fee of EUR 5 and a variable payment based on their performance in each of the two real-effort tasks. The average total payment was EUR 11.33. The total duration of the experiment was 45 minutes, resulting in average hourly earnings of EUR 15.11. (The laboratory target earnings rate was EUR 14/hour at the time of data collection.) The experimental interface was programmed in oTree \citep{Chen.2016}.  A total of 638 subjects participated in our experiments. See EC.1-EC.2 for the description of the experimental protocol, attention checks, stimuli and exclusions. The complete experimental instructions are available in an online depository (\url{https://researchbox.org/3723&PEER_REVIEW_passcode=PZXHPR}).

 %, leaving us with a total of 401 participants for that task. %There were at least 50 participants who completed each treatment and task. 

\section{Scrabble Task}
In this section we examine the effects of workflow in a Scrabble-based task. The open-ended, creative nature of Scrabble lends itself to studying the relevant behaviors and drivers of performance in a setting where the solution space is very large and ex-ante unknown, and participants must discover, explore and recombine various ideas to achieve good results. 

\subsection{Experimental Setup and Hypotheses}
\subsubsection{Setup}
At the start of the task, participants receive a set of tiles with letters on them. Words must be formed and connected in crossword fashion, and must read left to right or top to bottom. Deviating from the classic version of Scrabble, there are two separate boards that represent two product features or components. On one board subjects may only form nouns, and on the other board they may only form verbs.  Each board has $15\times15$ fields. For each board, subjects receive 100 letters with no refill. The list of letters is the same for each participant. Words cannot be formed diagonally. Each used letter is worth five points. Screen shots of the experimental interface are shown in Figure \ref{fig:Scrabble_screens} in the Appendix. The validity of each word placed on the board is instantly checked against the online dictionary wiktionary.org and highlighted in green if valid. The overall performance, used to determine participant compensation, is computed as the minimum of the two component scores (verb and noun scores).  This is to represent that a product has multiple components, and each of the components needs to be done well before the product can be taken to market. This payoff function also ensures that participants are incentivized to work on both tasks, instead of working on the task they consider to be easier or more enjoyable. In \textsection 7 we will examine an alternative payoff function  based on the sum of the two components. 

%The overall performance, used to determine participant compensation, is computed as follows. First, the number of letters is counted separately for each board. For overlapping words, the overlapping letter is counted twice. For example, in Figure \ref{fig:Scrabble_screens}, the noun task (panel a) has four words with 5, 6, 4 and 4 letters respectively. Subjects receive 5 points for each letter, yielding $(5+6+4 + 4) \times 5 = 95$ points in this example. The score is computed analogously in the verb task. For example, in Figure \ref{fig:Scrabble_screens}, the subject has five verbs with 4, 5, 6, 6 and 6 letters respectively, yielding the total score of $(4+5+6+6 + 6)\times 5 = 135$. At the end of the activity, the smaller of the two task scores becomes the final payoff. This is to represent that a product has multiple components, and each of the components needs to be done well before the product can be taken to market. In this example the participant would earn  the lower of 95 and 135, i.e., 95 points. This payoff function ensures that participants work on both tasks, instead of working on the task they consider to be easier or more enjoyable. %While other payoff functions may be equally suitable to represent complementarities between components, our choices were driven mainly to facilitate comprehension and easy calculation of profits for participants. 

\subsubsection{Treatments (Between-Subject)}
We consider four between-subject treatments. In all four treatments the overall time is 12 minutes, and there are two periods of equal length. The workflow, i.e., the allocation of time to tasks depends on the treatment. In the \textit{SEQ} treatment participants complete the task sequentially, with only one task being worked on in each period. In the \textit{ITER} treatment participants work on both tasks in parallel throughout the time horizon, switching between tasks as they see fit. In the \textit{ITER EQUAL FREEZE} treatment, the words placed on the board in the first period are ``frozen'' during the second period and cannot be (re-)moved. In addition, in this treatment participants must allocate exactly the same amount of time to each task. Thus, the only difference between \textit{SEQ} and \textit{ITER EQUAL FREEZE} is that participants in the latter treatment are allowed to task-switch. The \textit{ITER FREEZE} treatment is analogous, but the equal time allocation constraint is relaxed. The differences in workflow between the treatments are summarized in Table \ref{tab:scrab design}.  

\begin{table}[tb]
\TABLE
{Scrabble Treatments\label{tab:scrab design}}
{\small
  \centering
  \renewcommand{\arraystretch}{1.3}
    \begin{tabular}{lC{2.5cm}C{3.4cm}C{2.8cm}C{2.9cm}}
\toprule
    \textbf{Treatment}  & \textbf{Work periods} & \textbf{Task switching allowed within period?}	& \textbf{Time allocation to tasks flexible?}	& \textbf{Can period 1 work be altered during period 2?} \\
    \midrule
    \textit{SEQ}     & Two periods, \newline six minutes each & No   & No   & No \\
    \textit{ITER EQUAL FREEZE}   & Two periods, \newline six minutes each & Yes    & No & No \\
    \textit{ITER FREEZE}   & Two periods, \newline six minutes each & Yes   & Yes    & No \\
    \textit{ITER}     & Two periods, \newline six minutes each & Yes    & Yes   & Yes \\
    \bottomrule
    \end{tabular}%
}
{}
\end{table}

\subsubsection{Hypotheses}
The standard economic argument is that a less constrained action set should improve performance.  In our setting, \textit{SEQ} presents workers with the most constraints, while \textit{ITER} is the least-constrained workflow (See Table \ref{tab:scrab design}). We use this reasoning to formulate the following hypothesis  regarding the effects of workflow on performance.

\smallskip

\noindent \textit{\textbf{H1.1:} Average performance is higher in ITER than in SEQ.}

\noindent \textit{\textbf{H1.2:} Average performance is ranked as follows: SEQ $<$ ITER EQUAL FREEZE $<$ ITER FREEZE $<$ ITER.}

Counterarguments to H1.1-H1.2 are found in several behavioral studies showing that constraints can be helpful in some complex tasks \citep{lurie2009, sawyer2011, kagan2018, long2020}. Other research  found that when given a choice, workers overspend time and energy on the easier, rather than the most value-adding tasks \citep[][see \textsection 2.1 for details]{ibanez2018,kc2020}. While it is not clear whether these types of behaviors will occur in our setting, these studies suggest that certain types of constraints may indeed be beneficial for performance. 

%\subsubsection{Parametrization and Experimental Protocol}

%\subsection{Experiment Design} 
%We conducted pre-tests with 33 participants to calibrate the durations of tasks,  the materials (number of letters in the Scrabble task), and the payoff landscapes for the Lemonade Stand task (mapping between attributes and profit). Task durations and the number of letters available were chosen to ensure that both time and material constraints were binding for most participants, yet sufficient for some to achieve top performance. We found that these goals were achieved with 100 letters per component and 6 minutes per period in the Scrabble-based tasks, and 4 minutes per period in the Lemonade Stand task.  

\subsection{Results} 
\subsubsection{Summary Statistics} 
\begin{figure}[b]
\FIGURE
{\includegraphics[width=0.75\textwidth]{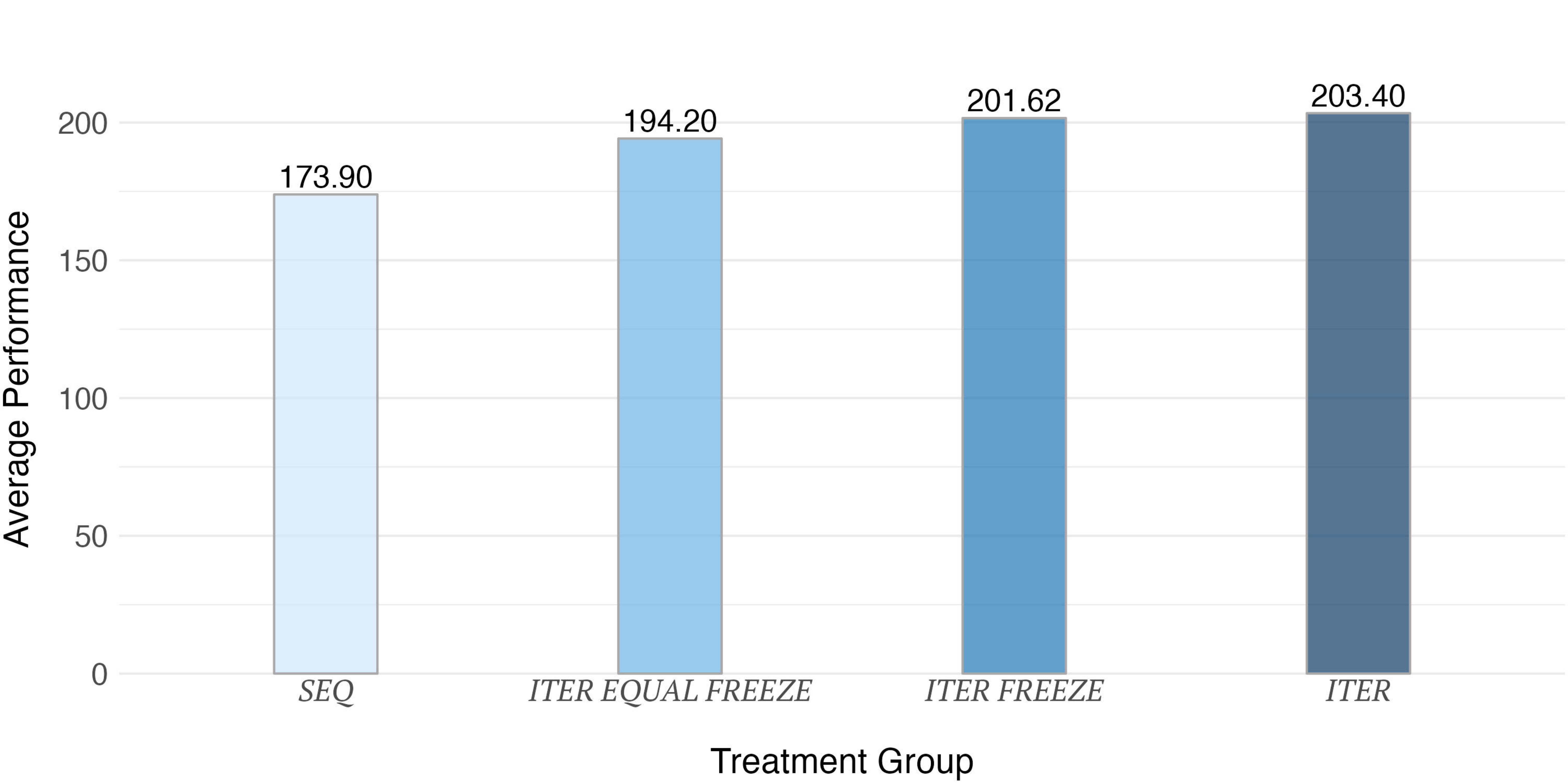}}
{Scrabble: Performance by Treatment\label{fig:s1_bars}}
{}
\end{figure}

Figure \ref{fig:s1_bars} shows the average treatment performance ($MIN\{Component~1$ $Score$,  $Component~2~Score\}$). The lowest performance is observed in the sequential workflow treatment (\textit{SEQ}). Further, the largest gap is between the \textit{SEQ} and the iterative treatments. Both of these patterns are in line with H1: fewer workflow constraints lead to better performance. However, some of the treatment differences are quite small. In particular, while the differences in mean performance range from  20.3 to 29.5 points between \textit{SEQ} and  iterative treatments, the differences within the iterative treatments are minimal (ranging between 1.8 and 9.8 points). This suggests that some of the H1 predictions may not be supported in the data, and that the key driver of performance differences is the ability to task-switch, rather than the presence (or absence) of additional time and process constraints.

\subsubsection{Hypothesis Tests} 
We next test H1 using regression analysis. Table \ref{tab:Reg1} shows the estimates of the treatment effects, with the \textit{SEQ} treatment indicator serving as the baseline for comparisons. Columns (1)-(2) show the effects on overall performance (participant payoff) - with and without demographic control variables. Column (1) presents raw treatment effects (without including any demographic or treatment controls), and shows significant effects of both \textit{ITER FREEZE} and \textit{ITER}. Column (2) adds session and demographic controls. This specification shows stronger performance effects of all iterative treatments, ranging from 24.41 to 46.39 points (between 14 and 28 percentage point improvement, computed based on marginal effects), with two of the three $p-$values below 0.01. 

In columns (3)-(4), we examine productivity gains in each period, i.e., the score improvements achieved in period 1 and in period 2 (Recall that in \textit{SEQ} each period corresponds to a single component, while in iterative treatments subjects can work on both components during each period). We find that the performance advantage of the iterative treatments is driven primarily by the poor performance of sequential workflow during the first period (column 3, $p\ll0.01$). In contrast, if we focus on the second period (column 4), the differences between treatments are smaller and not statistically significant ($p=0.196$). Finally, the bottom panel of Table \ref{tab:Reg1} shows that constraints improve productivity during the first period with \textit{ITER EQUAL FREEZE} outperforming both other iterative regimes ($p=0.008$ and $p=0.042$  in col. 3). However, the overall performance differences between different iterative regimes are not significant ($p>0.162$ for all pairwise tests in col. 1-2). Taken together, these analyses suggest that the ability to switch back-and-forth and multitask is a key performance differentiator between iterative and sequential workflows. In contrast, time and process constraints play a subordinate role. Therefore, H1.1 is supported, while H1.2 is not. 

\begin{table}[t]
\def\sym#1{\ifmmode^{#1}\else\(^{#1}\)\fi}
\TABLE
{Scrabble: Hypothesis Tests\label{tab:Reg1}}
{\small
  \centering
\begin{tabular}{l C{1.8cm} C{1.8cm} C{2cm} C{2cm}}
    \toprule
    & (1)   & (2)   & (3)   & (4)   \\
    \multicolumn{1}{r}{Dependent Variable:} & \textit{Performance} & \textit{Performance} & \textit{Period 1 Productivity} & \textit{Period 2 Productivity}    \\ 
    \midrule
    \textit{SEQ} & Baseline & Baseline & Baseline & Baseline \\ \addlinespace 
    \textit{ITER EQUAL FREEZE} & 20.29\sym{} & 46.19\sym{***} & 85.08\sym{***} & -20.00 \\
                     & (13.12) & (13.94) & (17.51) & (15.46) \\ \addlinespace
    \textit{ITER FREEZE}       & 27.71\sym{**} & 40.70\sym{***} & 40.97\sym{***} & -6.81 \\
                     & (12.60) & (12.29) & (15.44) & (13.63) \\ \addlinespace
    \textit{ITER}              & 29.50\sym{**} & 24.41\sym{*} & 45.22\sym{***} & -23.69 \\
                     & (13.56) & (12.93) & (16.25) & (14.35) \\\addlinespace
    Constant         & 173.9\sym{***} & 161.7\sym{***} & 184.7\sym{***} & 195.7\sym{***} \\
                     & (8.65) & (28.32) & (35.58) & (31.42) \\ \addlinespace
    Demographic and session controls         & No   & Yes   & Yes   & Yes   \\
    Observations     & 244   & 244   & 244   & 244    \\
    R-squared        & 0.027 & 0.182 & 0.168 & 0.204  \\
    \midrule
    \addlinespace
    \textbf{Pairwise tests} ($p-$values)       &  & &    &  \\  [1ex]
    \textit{ITER EQUAL FREEZE = ITER FREEZE} & 0.582 & 0.675 & 0.008 & 0.364 \\
    \textit{ITER EQUAL FREEZE = ITER}        & 0.523 & 0.162 & 0.042 & 0.830 \\
    \textit{ITER FREEZE = ITER}              & 0.898 & 0.240 & 0.807 & 0.273 \\
    \bottomrule
\end{tabular}%
}
{\textit{Notes.} OLS regressions with standard errors in parentheses. Columns (1)-(2) use overall performance, i.e., the minimum of the two component scores as the dependent variable. Columns (3)-(4) use period-wise productivity, i.e., the sum of score improvements achieved in each period.  Columns (2)-(4) control for age, gender, German native speaker, education,  task sequence, component sequence, loss of internet connection and familiarity with Scrabble. \sym{*}\(p<0.10\); \sym{**}\(p<0.05\); \sym{***}\(p<0.01\).}
\end{table}

\section{Scrabble with Pre-formed Words} 
The Scrabble task in \textsection 4 combines creative and analytical thinking, with good performance relying on successful idea generation, selection, and execution. This does not allow us to separately observe and measure how ideas are formed vs. how they are integrated with the existing ones (as many ideas are discarded internally before they appear on the board). Indeed, prior psychology literature has already identified that taking (even short) breaks from a creative task and switching to another activity may boost idea generation \citep[see][and references there]{sawyer2011}. In contrast, the effects of workflow and task-switching on idea selection and recombination activities have not been studied in the literature (see \textsection 2) and are less clear. To broaden our insights, in this section we will introduce a new experimental task that helps isolate these more downstream innovation behaviors from idea generation.

\subsection{Experimental Setup and Hypotheses}
The general experimental protocol (participants, payments, exclusions etc.) is described in \textsection 3 and EC.1. Here we focus on the key changes in the design relative to the Scrabble task  of  \textsection 4. 

\subsubsection{Task}
The new task retains the physical/spatial design component of the original Scrabble task of \textsection 4 requiring participants to combine elements in a 2D space, but restricts them from producing new ideas.  Instead, participants need to examine different combinations of existing ideas.  To this end, we use a smaller, 5$\times$5 Scrabble board and require participants to \textit{only} use combinations of 20 five-letter pre-formed words, either nouns or verbs (We made the board smaller relative to \textsection 4 to ensure that the task was sufficiently difficult). As before, the goal is to use as many letters as possible. We chose a single combination of words leading to a global optimum (six words, i.e., 30 letters, equivalent to 150 points) and two combinations that lead to the next best result (125 points).  Screenshots are in Figure~\ref{fig:study2_design_task_screenshots} in the Appendix.

We administered two between-subject treatments: \textit{SEQ} and \textit{ITER}. Similar to the previous version of the Scrabble task, in the \textit{SEQ} treatment participants were only allowed to work on a single task (verbs or nouns) in each of the two periods. However, in the \textit{ITER} treatment they were allowed to switch between tasks and work iteratively. Given the lack of notable differences among the different iterative workflow treatments in \textsection 4, we omit the intermediate \textit{ITER FREEZE} and \textit{ITER FREEZE EQUAL} treatments in this task. 

\subsubsection{Hypotheses}
The open-ended, creative nature of the Scrabble task means that there are several potential pathways for the treatment effects identified in \textsection 4. One possible pathway is that iterative workflow facilitates idea production and removes creative blocks -- a performance barrier that has been documented in the idea generation and brainstorming literature \citep[][and references there]{sawyer2011}. Working concurrently on two tasks (verbs and nouns) may benefit creative thinking and help unblock creative production, leading to the generation of a greater number of ideas. If this mechanism is the main driver of the treatment effects, then the performance advantage of iterative workflow should collapse once idea generation is muted. Furthermore, the disadvantages of iterative work, such as the cognitive costs of switching between tasks  \citep{gilovich2002,kiesel2010}, as well as potential increases in the cognitive load \citep[][see \textsection 2.1-2.2 for details]{lurie2009,kagan2018,long2020} may neutralize any benefits of increased flexibility. This leads us to the following hypothesis:

\noindent \textit{\textbf{H2:} When participants are provided with pre-existing ideas (words), average performance does not differ between \textit{SEQ} and \textit{ITER}. }

An alternative pathway for the benefits of iterative workflow is that it does not affect the generative stages of idea production, but rather the selection and integration of ideas. \cite{girotra2010} and \cite{kagan2018} highlight the importance of idea selection in creative performance, showing that certain time management strategies prevent inertia and improve selection quality. If this holds in our setting, H2 would be rejected and iterative workflow would continue to dominate.

\subsection{Results}
\subsubsection{Summary Statistics}
\begin{figure}[t]
\FIGURE
{\includegraphics[width=0.45\textwidth]{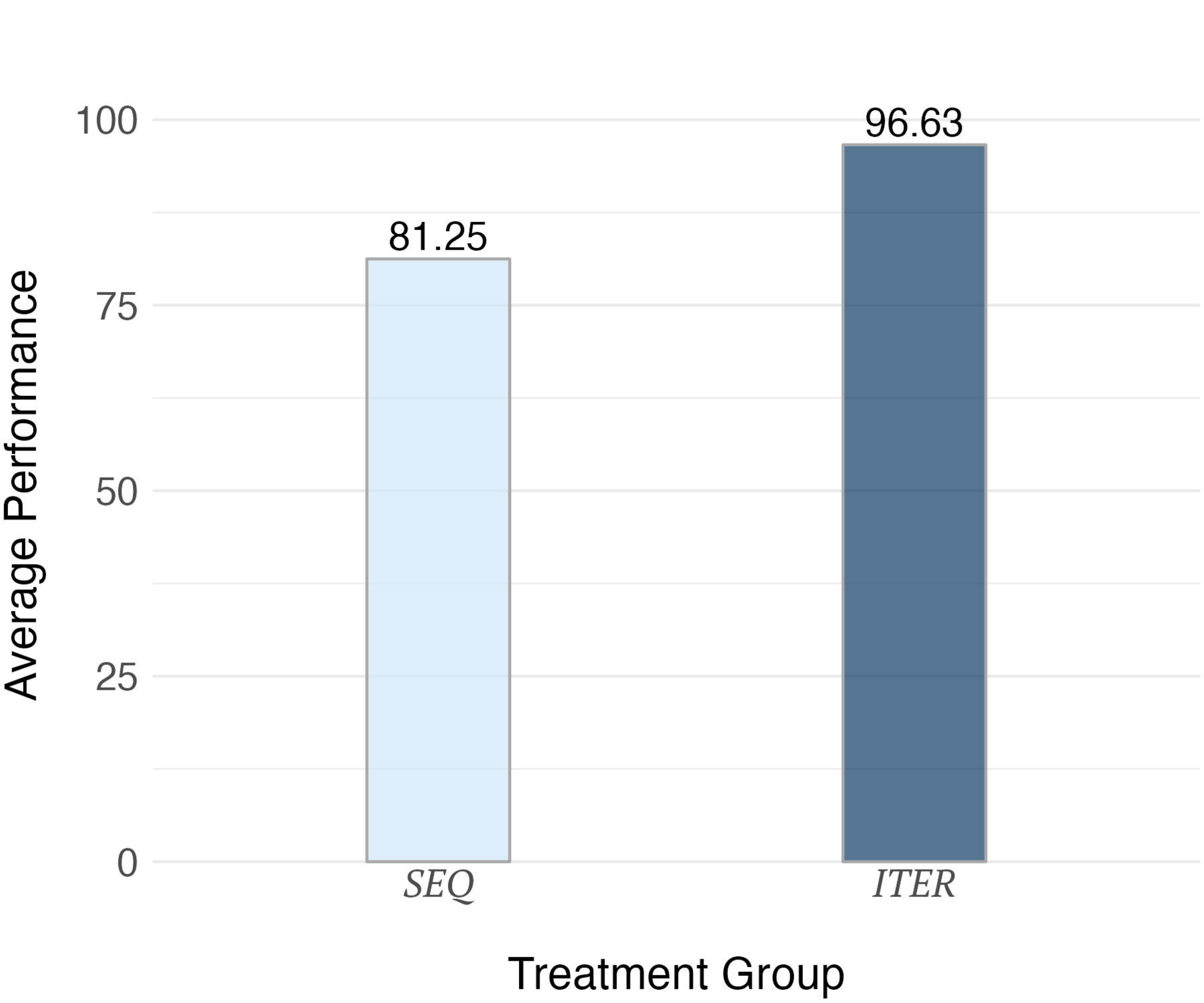}}
{Scrabble with Pre-formed Words: Performance by Treatment\label{fig:s2_bars}}
{}
\end{figure}

Figure \ref{fig:s2_bars} shows average treatment performance. Similar to the classic Scrabble task  in \textsection 4, there is a notable difference in treatment performance. Further, the magnitude of the difference is quite similar to the original Scrabble task: the improvement going from \textit{SEQ} to \textit{ITER} is 18.93\%. This provides preliminary evidence counter to H2, i.e., even when idea generation is muted,  iterative workflow continues to outperform sequential.

\subsubsection{Hypothesis Tests}
To test H2 more formally, we conduct a series of regressions. The regression coefficients are reported in Table \ref{tab:Reg2}. The regression results do not support H2: \textit{ITER} workflow continues to dominate both in the absence of demographic and session controls ($p=0.011$, column~1), and after including these controls ($p=0.019$, column 2). As before, the treatment difference is driven primarily by the quick productivity gains made by participants in the \textit{ITER} treatment during the first work period ($\beta=80.36, p\ll0.001$, column 3). This is unsurprising given that the \textit{ITER} workflow allows participants to make progress on both components in each period. Equally unsurprising is that the \textit{SEQ} workflow outperforms \textit{ITER} during the second period ($\beta=-64.27, p\ll0.01$, column~4), as \textit{SEQ} starts to work on a new component in period 2 and can thus make gains at a faster rate. However, the difference in the sizes of the coefficients in column~3 and column~4 (80.36 vs. 64.27) suggests that \textit{SEQ} is not able to fully catch up with \textit{ITER}.  

These results suggest that the performance patterns observed in \textsection 4 replicate with the new task introduced in this section. Given that we muted idea generation and provided participants with existing ``ideas'' (words), idea generation cannot be the sole driver of our results. Instead, the ability to recognize effective combinations on a complex solution space of ideas appears to be an important pathway through which iterative workflow improves performance.

\begin{table}[tb!]
\def\sym#1{\ifmmode^{#1}\else\(^{#1}\)\fi}
\TABLE
{Scrabble with Pre-formed Words: Hypothesis Tests\label{tab:Reg2}}
{\small
  \centering
\begin{tabular}{l C{1.6cm} C{1.6cm} C{1.7cm} C{1.7cm}}
    \toprule
    & (1)   & (2)   & (3)   & (4)   \\
    \multicolumn{1}{r}{$~~~~~~~~~~~~~~~~~~$Dependent Variable:} & \textit{Performance} & \textit{Performance} & \textit{Period 1 Productivity} & \textit{Period 2 Productivity}    \\
    \midrule
    \textit{SEQ} & Baseline & Baseline & Baseline & Baseline \\ \addlinespace 
    \textit{ITER} & \multicolumn{1}{c}{15.38**} & \multicolumn{1}{c}{13.65**} & \multicolumn{1}{c}{80.36***} & \multicolumn{1}{c}{-64.27***} \\
          & \multicolumn{1}{c}{(5.98)} & \multicolumn{1}{c}{(5.71)} & \multicolumn{1}{c}{(7.97)} & \multicolumn{1}{c}{(6.34)} \\ \addlinespace 
    Constant & \multicolumn{1}{c}{81.25***} & \multicolumn{1}{c}{125.80***} & \multicolumn{1}{c}{130.50***} & \multicolumn{1}{c}{125.79***} \\
          & \multicolumn{1}{c}{(4.07)} & \multicolumn{1}{c}{(16.35)} & \multicolumn{1}{c}{(22.81)} & \multicolumn{1}{c}{(18.15)} \\  \addlinespace 
              Demographic and session controls         & No   & Yes   & Yes   & Yes   \\
    Observations & \multicolumn{1}{c}{112} & \multicolumn{1}{c}{112} & \multicolumn{1}{c}{112} & \multicolumn{1}{c}{112} \\
    R-squared & \multicolumn{1}{c}{0.057} & \multicolumn{1}{c}{0.234} & \multicolumn{1}{c}{0.558} & \multicolumn{1}{c}{0.517} \\
\bottomrule
\end{tabular}%
}
{\textit{Notes.} OLS regressions with standard errors in parentheses. Columns (1)-(2) use overall performance, i.e., the minimum of the two component scores as the dependent variable. Columns (2)-(4) use period-wise productivity, i.e., the sum of score improvements achieved in each period.  Columns (2)-(4) control for age, gender, German native speaker, education, task sequence, component sequence, loss of internet connection and familiarity with Scrabble. \sym{*}\(p<0.10\); \sym{**}\(p<0.05\); \sym{***}\(p<0.01\).}
\end{table}

\subsubsection{Detailed Analysis}
A key behavior characterizing the participant's approach is the number of times the participant removes a word from the board. This is because the action space is constrained by the small size of the board, so that performance improvements are only possible through repeated removal and repositioning of words. Examining the frequency of word additions and word removals, we find that on average, both are significantly higher with iterative flow (Additions: 9.73 times in \textit{ITER} vs. 11.81 times in \textit{SEQ}, $p=0.001$; Removals: 3.48 times in \textit{ITER} vs. 2.26 times in \textit{SEQ}, $p=0.024$). Further, the frequency of both word additions and word removals was strongly correlated with performance (Pearson correlation coefficient for additions and performance: $\rho=0.71$, $p\ll0.001$; for removals: $\rho=0.33$, $p=0.001$).  

Together, these comparisons indicate that the advantage of iterative workflow extends to innovation tasks that do not require idea generation but are based solely on the recombination of existing ideas. The \textit{ITER} treatment's pattern of frequent word removals further suggests that task-switching may prompt a ``fresh start effect,'' enabling participants to break from existing solutions and explore new combinations. In contrast, sequential workflow appears to encourage more linear, myopic behaviors. In the next section, we perform a sharper test of this mechanism and examine how explore-exploit patterns differ between iterative and sequential workflows.

\section{Lemonade Stand Task} 
In this section, we introduce a new experimental task that allows more precise control over both the solution landscape and the behaviors that drive innovative performance. This task will enable us to perform sharper tests of the proposed mechanisms and will help us identify several boundary conditions on the benefits of iterative workflow. 

\subsection{Experimental Setup and Hypotheses}
\subsubsection{Lemonade Stand Task}
Representing innovation activities as search on multidimensional landscapes is a common approach in the economics and management research  \citep{levinthal1981,levinthal1997,mihm2003,sommer2004,billinger2014}. As is common in the experimental implementation of landscape problems \citep[see, for example,][]{ederer2013, sommer2020} we use the naturalistic framing of designing and managing a "Lemonade stand" to represent the solution landscape.  The participant is asked to identify an effective business strategy by repeatedly choosing the values of several business attributes, and learning about the  payoff resulting from each attribute combination.  Deviating from the classic version of the Lemonade Stand game, we introduce two separate independent landscapes: one for  product and one for market attributes. The product landscape consists of four product attributes: lemonade color,  lemon content,  carbonation,  bottle label.   The market landscape also consists of four market attributes: location, price,  opening hours,  advertising.  Figure~\ref{fig:Lemo_screens} shows the decision screens for each of the two tasks. Participants can modify the attributes as often as they like. However, each time they do so, there is a 3-second delay until they see the resulting profit. This is to encourage thoughtful choices and to discourage random clicking. As before, participants were paid based on the lower of the two component scores.

\subsubsection{Treatments}
We examined two different versions (parametrizations) of the Lemonade Stand game: a \textit{rugged} landscape parametrization, and a \textit{smooth} one. These parametrizations were chosen and calibrated based on previous implementations of the Lemonade Stand game \citep[for example,][]{ederer2013,sommer2020}. For landscape visualizations please see EC.2.  In all parametrizations there is a single global optimum with a payoff of 500 points, and two further local optima, with payoffs of 200 and 380 points, respectively. In the rugged parametrization, discovering the global optimum requires more experimentation. This is because the locations of each of the three optima are different in each of the nine combinations of the discrete attributes. In contrast, in the smooth parametrization,  ``greedy'' myopic refinement can suffice to discover the global optimum. This is because the mapping between attributes and performance is less complex, with each attribute having a single, unique optimal value that does not depend on other attributes. For example, in the smooth  parametrization, it is always optimal to choose a price of 17.9 units, regardless of the other attributes. In contrast, in the rugged  parametrization, the optimal price depends on the choice of lemonade color and bottle label. As noted in \textsection 3 (Table \ref{tab:design}), we conduct four treatments (\textit{SEQ},  \textit{ITER}, \textit{ITER EQUAL FREEZE} and \textit{ITER FREEZE}) in the rugged landscape parametrization, and two treatments (\textit{SEQ} and \textit{ITER}) in the smooth parametrization.  To keep the analysis focused, in the main text, we only report the analysis related to the (\textit{SEQ} and \textit{ITER}) treatments. The \textit{ITER EQUAL FREEZE} and \textit{ITER FREEZE} treatments are discussed in EC.3, with only the key results summarized in the main text.\footnote{To impose freezing constraints in the Lemonade Stand game (\textit{ITER EQUAL FREEZE} and \textit{ITER FREEZE} treatments), we fix two of the four attributes in each component to their best discovered values after the first period. Participants are informed about which attributes will be frozen.}

\subsubsection{Hypotheses} To develop hypotheses we return to the theoretical discussions in \textsection 2.1-2.2 and in \textsection 4.1.3. Having more flexibility to choose how to allocate time and being able to iterate should generally improve performance. Based on our previous results,  the hypothesized pathway through which iterative workflow may facilitate performance improvements is task-switching. By moving between tasks, participants split the work into smaller increments, with each increment presenting an opportunity for a fresh start. In the context of the Lemonade Stand game, such opportunities are especially useful when the solution landscape is more rugged -- in this case restarting the search may prompt participants to experiment with more diverse combinations of attributes and can thus help perform a broader exploration of the landscape. In contrast, workflow should have minimal effects on smoother landscapes, where myopic, ``greedy'' optimizing of individual features can lead to good results. We formalize this logic as follows:

\noindent \textit{\textbf{H3.1:} On a rugged landscape, average performance in \textit{ITER} is better than in \textit{SEQ}.}

\noindent \textit{\textbf{H3.2:} On a smooth landscape, average performance does not differ between  \textit{ITER} and \textit{SEQ}.}

\subsection{Results}
\subsubsection{Summary Statistics}
\begin{figure}[bt]
\FIGURE
{\includegraphics[width=0.8\textwidth]{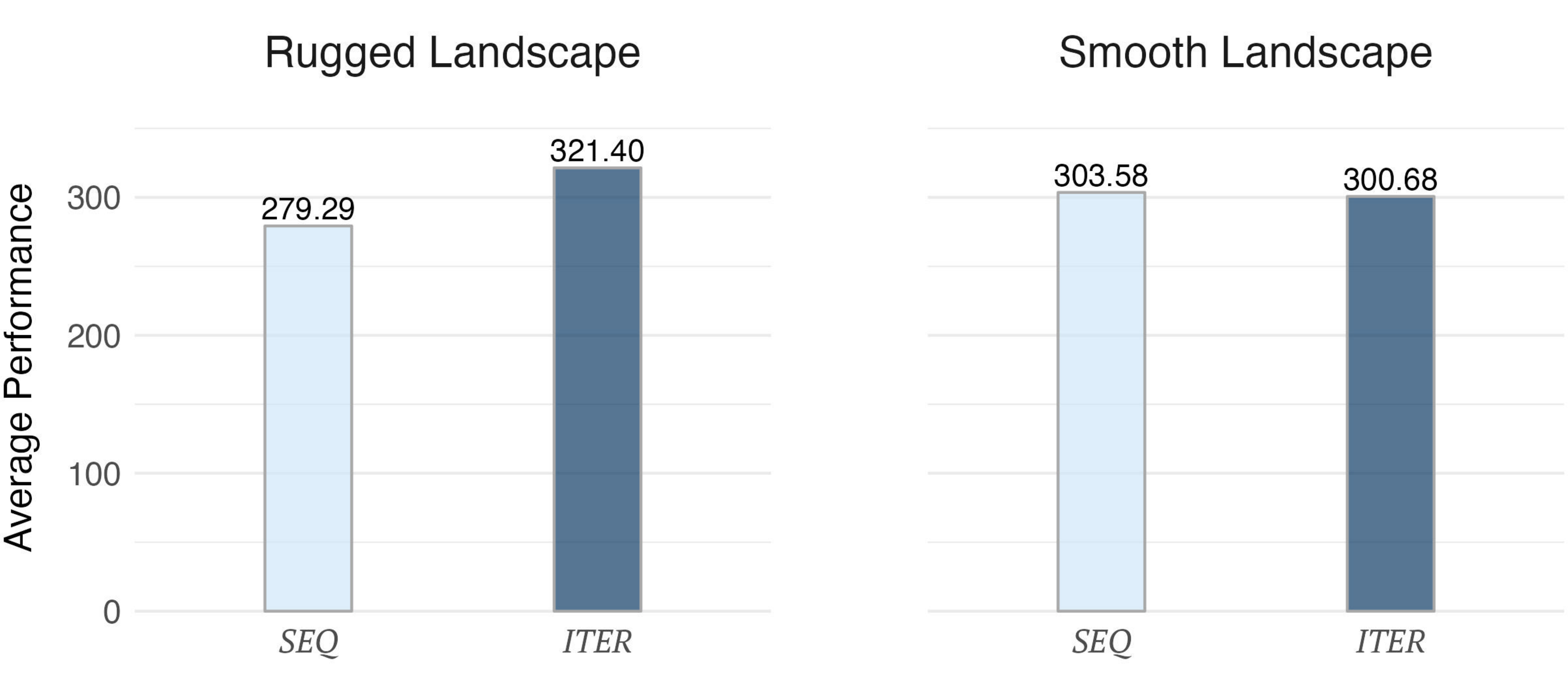}}
{Lemonade Stand: Performance by Treatment\label{fig:s3_bars}}
{}
\end{figure}

Figure \ref{fig:s3_bars} shows average performance in each treatment group. The figure suggests the following. First, the left part of the figure shows that \textit{ITER} outperforms \textit{SEQ} in the rugged landscape parametrization, with a performance improvement of 42.11 points. Second, the right part of the figure suggests only minimal effects of workflow in the smooth landscape parametrization (2.90 point difference between treatments). These comparisons suggest that both H3.1 and H3.2 are supported in our data. We next test these hypotheses formally using regression analysis. 

\subsubsection{Hypothesis Tests}
Table \ref{tab:RegLemo} shows regression coefficients in the rugged (columns 1-4) and smooth  (columns 5-8) landscape versions of the Lemonade Stand game. The results confirm that the treatment differences in performance observed in the left part of Figure \ref{fig:s3_bars} (rugged landscape) are significant. In particular, the treatment effect of \textit{ITER} is between 42.11 and 48.59 points, corresponding to an improvement of 15 to 18 percent relative to \textit{SEQ} ($p=0.009$ and $p=0.018$).  Further, column (3) shows that the benefits of iterative workflow, again, are caused by the low productivity of \textit{SEQ} in the first phase, with the treatment effect of \textit{ITER} being large (270.9 points) and significant at $p\ll0.001$. This is not surprising given that participants in \textit{SEQ} only have access to one of the two components. Given that performance has a sharply  diminishing pattern in this task, being restricted to working on a single component greatly reduces productivity in period 1. Equally unsurprising is the reversal of the performance gains in the second phase, as participants in \textit{SEQ} begin to work on the second component ($\beta=-225.2, p\ll0.001$ in column~4). However, the difference in coefficient sizes between column (3) and (4) suggests that the performance gains made in \textit{ITER} during the first phase are not fully offset by the performance gains in  \textit{SEQ} during the second phase. 

Columns (5)-(6) show that the treatment differences in overall performance are not statistically significant in the smooth landscape parametrization (lowest $p=0.187$). As before, we observe strong and statistically significant advantages of the \textit{ITER} workflow in the first phase (column 7), and of the \textit{SEQ} workflow in the second phase (column 8). However, this time, the effect size is greater in the second phase, resulting in the \textit{SEQ} treatment surpassing \textit{ITER} in terms of its productivity, which in turn explains the absence of a detectable treatment effect on overall performance observed in columns (5)-(6). 

\begin{table}[th]
\def\sym#1{\ifmmode^{#1}\else\(^{#1}\)\fi}
\TABLE
{Lemonade Stand Task: Regression Results\label{tab:RegLemo}}
{\small
  \centering
\begin{tabular}{l C{1.35cm}C{1.35cm} C{1.65cm}C{1.65cm}  c   C{1.35cm}C{1.35cm}C{1.65cm}C{1.65cm}}
\toprule
& \multicolumn{4}{c}{Rugged landscape} & & \multicolumn{4}{c}{Smooth landscape} \\
\cline{2-5} \cline{7-10}  \addlinespace
    & (1)   & (2)   & (3)   & (4)  &&    (5)   & (6)   & (7)   & (8)  \\
\multicolumn{1}{r}{Dep. Var.} & \textit{Performance} & \textit{Performance} & \textit{Period 1 Productivity} & \textit{Period 2 Productivity} &  & \textit{Performance} & \textit{Performance} & \textit{Period 1 Productivity} & \textit{Period 2 Productivity} \\ [0.5ex]
\cline{1-5} \cline{7-10} \addlinespace
\textit{SEQ} & Baseline & Baseline  & Baseline  & Baseline &  & Baseline &Baseline  &Baseline  &Baseline  \\ \addlinespace
\textit{ITER} & 42.11** & 48.59*** & 270.90*** & -225.20*** & & -2.89 & -5.45 & 176.70*** & -205.60*** \\
& (17.60) & (18.43) & (23.46) & (20.26) & & (17.95) & (18.42) & (23.04) & (23.72) \\
\addlinespace
Constant & 279.3*** & 283.4*** & 308.0*** & 399.7*** & & 303.6*** & 385.8*** & 317.6*** & 533.2*** \\
& (10.45) & (48.66) & (61.94) & (53.50) & & (12.42) & (52.68) & (65.90) & (67.84) \\
\addlinespace
Dem. and session controls         & No   & Yes   & Yes   & Yes & & No   & Yes   & Yes   & Yes  \\
Observations & 142 & 142 & 142 & 142 & & 115 & 115 & 115 & 115 \\
R-squared & 0.039 & 0.080 & 0.512 & 0.544 & & 0.000 & 0.071 & 0.379 & 0.464 \\
\bottomrule
\end{tabular}
}
{\textit{Notes.} OLS regressions with standard errors in parentheses. Columns (1)-(2) and (5)-(6) use overall performance, i.e., the minimum of the two component scores as the dependent variable. Columns (3)-(4) and (7)-(8) use period-wise productivity, i.e., the sum of score improvements achieved in each period.   Controls are age, gender,  education, parameter version, task, German native speaker and component sequence. \sym{*}\(p<0.10\); \sym{**}\(p<0.05\); \sym{***}\(p<0.01\).}
\end{table}

\subsubsection{Detailed Analysis}
In the Lemonade Stand task, performance is typically clustered in the vicinity of local optima \citep{ederer2013}. In Table \ref{tab:lemo_optima} we summarize the share of participants reaching each optimality region. The top row of the table shows that the share of participants reaching the global optimum is similar across treatments (Proportion test, lowest $p=0.53$). However, there are some treatment differences for the remaining regions. In particular, the proportion of participants in the middle region increases from 28.3\% in \textit{SEQ} to 56.0\% in \textit{ITER} in the rugged parametrization and from 46.7\% in \textit{SEQ} to 60.0\% in \textit{ITER} in the smooth parametrization (both $p<0.01$). Therefore, iterative workflow appears to mainly help low-to-medium performers improve their performance.  

\begin{table}[h!]
\def\sym#1{\ifmmode^{#1}\else\(^{#1}\)\fi}
\TABLE
{Lemonade Stand Task: \% of Subjects in Each Optimality Region\label{tab:lemo_optima}}
{\small
  \centering
    \begin{tabular}{lC{1.3cm}C{1.3cm} cC{1.3cm}C{1.3cm}}
    \toprule
          & \multicolumn{2}{c}{\textbf{Rugged Parametrization}} &       & \multicolumn{2}{c}{\textbf{Smooth Parametrization}} \\
          & \textit{SEQ}&   \textit{ITER} &       & \textit{SEQ} & \textit{ITER} \\
\cmidrule{2-3}\cmidrule{5-6}       \% of subjects reaching \textbf{global} optimum region & 19.6\%   & 24.0\% &       & 18.3\% & 14.5\% \\
    \% of subjects reaching \textbf{middle} local optimum region & 28.3\%   & 56.0\% &       & 46.7\% & 60.0\% \\
     \% of subjects reaching \textbf{low} local optimum region & 52.2\%   & 20.0\% &       & 35.0\% & 25.5\% \\
\cmidrule{2-3}\cmidrule{5-6}   Total & 100.0\% & 100.0\%   &       & 100.0\% & 100.0\% \\
\bottomrule
\end{tabular}
}
{}
\end{table}

In EC.3 we present the results of two additional treatments involving freezing (in the rugged parametrization): \textit{ITER FREEZE} and \textit{ITER EQUAL FREEZE}. We find that freezing significantly reduces performance compared to both \textit{SEQ} and \textit{ITER} conditions ($p-$values between 0.001 and 0.049). This result aligns with intuition. Recall that freezing was implemented by fixing two of four attributes during the second period to their optimal values from the first period. We observe that such constraints significantly affect performance, as fewer subjects are able to discover the global optimum region. Thus, different from the Scrabble task where freezing did not appear to be a binding constraint, in the more combinatorial  setting of the Lemonade Stand task, freezing can have significant effects on performance.  

\section{Incentives, Outcome Measures, Learning and Mechanisms}
This section introduces in-depth analyses that help deepen our understanding of the main results. First, in \textsection 7.1 we examine the effects of an alternative incentive system, which incentivizes participants based on the sum (as opposed to the lower) of the two component scores. Second, in \textsection 7.2 we look at alternative outcome measures. Third, in \textsection 7.3, we examine learning and show an effective lever for reducing the negative performance effects of sequential workflow. Finally, in \textsection 7.4 we unpack a key mechanism driving the advantage of the iterative workflow in the Lemonade Stand task. 

To preview the results of these analyses, we find the following. Differences in incentives have only minimal effects on performance, with iterative workflow continuing to outperform sequential workflow regardless of the incentive system used. However, the choice of the outcome (performance) measure is important. The workflow effects are only significant when we use the lower of the two performance scores as our outcome measure, and are not significant for the higher of the two component scores, or their sum. This is explained by the lower performance of sequential workflow in the initial component. Finally, we show that the performance advantage of iterative workflow in the Lemonade Stand task is tied to broader exploration of the solution space, particularly early on in the task. 

\subsection{Alternative Incentive Treatments}
\subsubsection{Tasks and Treatments}
To ensure that our effects were not caused by the specific incentive structure (compensating for the lower of the two component scores), we conducted two additional treatments: \textit{SEQ ALT} and \textit{ITER ALT}.  In both treatments, participants completed the Scrabble task (from \textsection 4) and the rugged version of the Lemonade Stand task (from \textsection 6).\footnote{We did not examine the effects of incentives in the smooth version of the lemonade task given that it showed no significant treatment effects (See \textsection 6 for details).}  Similar to the original \textit{SEQ} treatment of the Scrabble task, in the \textit{SEQ ALT} treatment participants were only allowed to work on a single component (verbs or nouns) in each of the two periods. Similarly, in the Lemonade Stand task, participants could only work on the market or product landscape during each period.  As before, in the \textit{ITER ALT} treatment subjects were allowed to switch between components and work iteratively and spend as much time as they want in each component. The only difference between the original (\textit{ITER} and \textit{SEQ}) and the new (\textit{ITER ALT} and \textit{SEQ ALT}) treatments was the compensation system. In particular, in the \textit{ALT} treatments we used the sum of scores (as opposed to the lower of the scores in the original treatments) to compensate participants. The experimental protocol, subject pool and exclusion criteria were unchanged relative to the previous experiments and are described in \textsection 3 and EC.1. 

\subsubsection{Results}
For brevity, we only report the results pertaining to potential differences in the effects of incentives (Detailed regression estimates are in EC.3). Our main variables of interest are the main effects of incentive system (minimum-based vs. sum-based), as well as potential interaction effects between workflow type (sequential or iterative) and incentive system (minimum-based vs. sum-based), which would indicate that the effect of workflow depends on how participants are compensated. We find no significant main effects of incentives: the $p-$values for overall performance was 0.580 for the Scrabble task and 0.752 for the Lemonade Stand task. Moreover, we find no significant interactions in either task: the $p-$values for overall performance was 0.682 for the Scrabble task and 0.129 for the Lemonade Stand task. These null results suggest that the structure of incentives does not play a central role in our tasks. After controlling for incentive system, iterative workflow continued to lead to significantly better overall performance, with a 23.91-point advantage ($p=0.047$) in the Scrabble task and a 28.54-point advantage ($p=0.018$) in the Lemonade Stand task. These results provide strong evidence that the performance advantages of iterative workflow are not artifacts of specific incentives used, but rather reflect fundamental differences in how people approach innovation tasks under different workflow constraints.\footnote{Additionally, we examine the effects of workflow in the subsample of participants in \textit{SEQ ALT} and \textit{ITER ALT}. While we do not have enough power to detect statistical differences in this subsample ($N=153$ in the Scrabble task, $N=154$ in the Lemonade Stand task), the effects are directionally correct and their size is similar to the ones reported in Table 7.}

\subsection{Alternative Outcome Measures}
We have so far focused on the lower of component scores as our outcome measure. This favors balanced development across both components, as performance is constrained by the weaker component. However, some firms may be more interested in the sum of component scores, i.e., total capability across dimensions, or even the maximum of component scores, relevant for firms seeking to differentiate along a single dimension, such as startups pursuing a focused competitive advantage. To examine the effects of workflow on these alternative outcome measures, in Table \ref{tab:RegResultsAltMetrics} we use the pooled sample of iterative and sequential treatments (\textit{SEQ, ITER, SEQ ALT} and \textit{ITER ALT}) and examine treatment effects on the minimum of component scores (columns 1 and 4), maximum of component scores (columns 2 and 5) and sum of component scores (columns 3 and 6). The table shows that iterative search yields significant performance improvements only when performance is measured as the minimum of component scores. Specifically, the \textit{ITER} treatment effect is positive and significant for the minimum metric in both tasks (18.90 for Scrabble, $p<0.05$; 26.99 for Lemonade Stand, $p<0.05$), but insignificant for both the maximum and sum metrics across both tasks (both $p>0.1$). This suggests that the benefits of iterative search emerge specifically through balanced improvements across components rather than through excelling in any single dimension or through aggregate performance gains. To better understand this result, in the next subsection, we examine how performance gains accrue over time.\footnote{Table \ref{tab:RegResultsAltMetrics} also suggests that the incentive system itself can play a role for alternative incentive measures. In particular, in the Scrabble task with sum-based incentives, the maximum of the component scores goes up significantly (col. 2), and the sum of scores goes up directionally but not significantly (col. 3). While these incentive effects align with intuition, they are still worth emphasizing.}

\begin{table}[tbh]
\def\sym#1{\ifmmode^{#1}\else\(^{#1}\)\fi}
\TABLE
{Regression Results for Alternative Performance Metrics\label{tab:RegResultsAltMetrics}}
{\small
  \centering
\begin{tabular}{l C{2.cm}C{2cm}C{1.8cm} c C{2cm}C{2cm}C{1.8cm}}
\toprule
& \multicolumn{3}{c}{Scrabble Task} & & \multicolumn{3}{c}{Lemonade Stand (Rugged Landscape) Task} \\
\cline{2-4} \cline{6-8}  \addlinespace
    & (1)   & (2)   & (3) &&    (4)   & (5)   & (6)  \\
\multicolumn{1}{r}{Dep. Var.} & \textit{$MIN\{$Comp. 1 Score, Comp. 2 Score$\}$} & \textit{$MAX\{$Comp. 1 Score, Comp. 2 Score$\}$} & \textit{Sum of Component Scores} &  & \textit{$MIN\{$Comp. 1 Score, Comp. 2 Score$\}$} & \textit{$MAX\{$Comp. 1 Score, Comp. 2 Score$\}$} & \textit{Sum of Component Scores} \\ [0.5ex]
\cline{1-4} \cline{6-8} \addlinespace
        \textit{SEQ} & Baseline & Baseline & Baseline & & Baseline & Baseline  & Baseline \\ \addlinespace
\textit{ITER} & 18.90** & 2.875 & 21.77 &  & 26.99** & -7.771 & 19.41 \\
& (9.513) & (8.756) & (16.37) & & (12.33) & (11.95) & (21.16) \\
\addlinespace
\textit{Sum-based Incentive} & 4.83 & 21.95** & 26.78 & & -5.417 & 1.778 & -3.874 \\
 \textit{System (ALT)}& (9.615) & (8.849) & (16.55) & & (12.16) & (11.71) & (20.74) \\
\addlinespace
Constant & 187.6*** & 226.2*** & 413.8*** & & 277.3*** & 444.9*** & 728.2*** \\
& (29.63) & (27.27) & (51.0) & & (14.3) & (37.77) & (66.88) \\
\addlinespace
Dem. and session controls         & Yes   & Yes   & Yes & & Yes   & Yes   & Yes  \\
Observations & 276 & 276 & 276 & & 296 & 296 & 296 \\
R-squared & 0.145 & 0.149 & 0.162 & & 0.019 & 0.064 & 0.022 \\
\bottomrule
\end{tabular}
}
{\textit{Notes.} OLS regressions with standard errors in parentheses. Demographic controls are age, gender, German native speaker, education. \sym{*}\(p<0.10\); \sym{**}\(p<0.05\); \sym{***}\(p<0.01\).}
\end{table}

\subsection{Learning and Task Sequencing in Sequential Workflow} 
Our findings in \textsection 4-6 indicate strong productivity effects of iterative (relative to sequential) workflow during the \textit{initial} work period. However, we observe no such effects (or even reverse effects) in the second period. To better understand this result, we can leverage the fact that the order of components in sequential workflow treatments was randomized in our experiment; we can therefore examine performance conditional on the specific sequence of components encountered. As before, we focus on the Scrabble task and on the rugged version of the Lemonade Stand task. Further, to increase statistical power, we pool the \textit{SEQ} and \textit{SEQ ALT} treatment data and \textit{ITER} and \textit{ITER ALT} treatment data. 

\begin{figure}[bt!]
\FIGURE
{\includegraphics[width=1.1\textwidth]{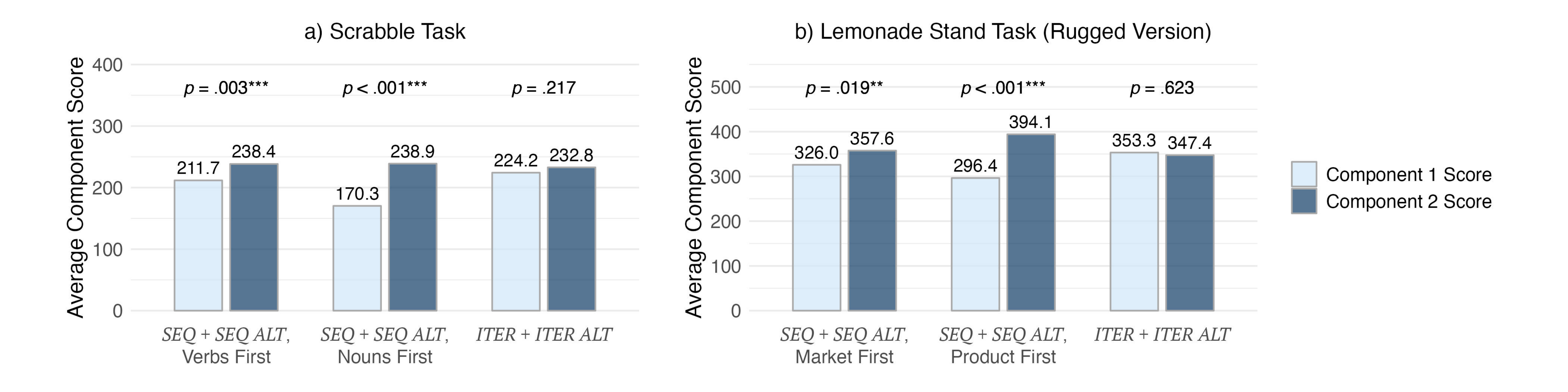}}
{Learning and Task Sequence (Component-wise Performance)\label{fig:learning}}
{$p-$values are based on within-subject signed rank test comparisons. $^*p<0.10; ~^{**}p<0.05; ~^{***}p<0.01$.}
\end{figure}

Figure \ref{fig:learning} shows performance in the first and second displayed components across two tasks under sequential and iterative treatments. Panel (a) shows results for Scrabble, while Panel (b) shows results for the Lemonade Stand (Rugged Version) task. As can be seen, both tasks show strong learning effects, with participants in sequential workflow consistently improving from the first to the second component they encounter. In the Scrabble Task (Panel a), performance on the second component significantly exceeds the first component performance,  regardless of presentation order ($p = 0.003$ for Verbs First; $p < 0.001$ for Nouns First). Similarly, the Lemonade Stand task (Panel b) shows the same pattern of improvement from first to second component ($p = 0.019$ for Market First; $p < 0.001$ for Product First). Notably, when participants begin with the more difficult component (Verbs in the Scrabble task or the Product component in the Lemonade Stand) they achieve lower initial performance compared to when they start with the easier component (Nouns or Market component, respectively). As we have seen previously, this initial slowdown has negative effects on subjects' overall performance, measured as the minimum of Component 1 and Component 2 scores. This pattern is remarkably consistent across both tasks, suggesting that the learning mechanisms are robust across tasks. In contrast, there is relatively little change in component-wise performance in the iterative treatments ($p = 0.217$ for Scrabble; $p = 0.623$ for Lemonade Stand). 

In addition to characterizing the performance changes over time, Figure \ref{fig:learning} suggests an important lever for improving the performance of sequential workflow: completing the easier component first to ensure overall acceptable performance across all components. Indeed, formal analysis shows that the sequence of components (easy $\longrightarrow$ difficult) has a substantial effect on the overall performance of sequential workflow. In Appx EC.3 we show that, for the subset of subjects who experienced the easier component first, the effects of workflow, while still directionally consistent, are no longer statistically significant ($p>0.1$). In contrast, the effects of workflow remain statistically significant (at $p<0.05$ for both Scrabble and Lemonade Stand), if we focus on the subjects that start with the more difficult component. These results suggest that in settings where iterative workflow may not be implementable, workers (and their organizations) can benefit from reordering the sequence of components from easiest to most difficult. A broader take-away of this analysis is that the benefits of iterative workflow are stronger for novel tasks where workers lack prior experience, and are less relevant for tasks that are routine or well-practiced.

\subsection{Exploration Behavior in Lemonade Stand}
Recall that we have introduced the Lemonade Stand because it allows us to more precisely examine search behavior and to compare how this behavior differs by treatment. In particular, to better understand the performance advantage of iterative workflow we next focus on exploration behavior, i.e., the extent to which participants explore a diverse vs. a narrow set of possibilities on the solution landscape. 

We quantify exploration behavior using the Herfindahl-Hirschman Index (HHI). To determine HHI, we first measure how often the same level of an attribute appears in a participant's solutions. Each of the two continuous attributes (lemon content and carbonation in the product landscape and price and opening hours in the market landscape) can take values from the range $[10, 10.1,...,19.9, 20]$, resulting in 101 levels per attribute. Each of the two discrete attributes (color and label in the product landscape and location and advertisement in the market landscape) comes with three levels. To calculate HHI, we count how often each level has been selected by the participant. We then compute a participant's HHI as the sum of the squared level selections relative to the total number of selected levels by the participant. Since each solution requires the selection of exactly four levels (one for each attribute), we multiply the HHI by four, such that if a participant always selects the same four levels for the same attributes, the HHI reaches its maximum value of one. Consequently, a larger HHI indicates more narrow, hill-climbing search behavior as the solutions explored by the participants are more similar to each other, whereas a smaller HHI value indicates a less concentrated, more exploratory search behavior. For a more intuitive presentation, we will use the complement of HHI, i.e, 1$-$HHI, which describes the dispersion (as opposed to the concentration) of explored solutions. We will use the term ``exploration behavior'' when referring to this measure.

\begin{figure}[bt!]
\FIGURE
{\includegraphics[width=0.9\textwidth]{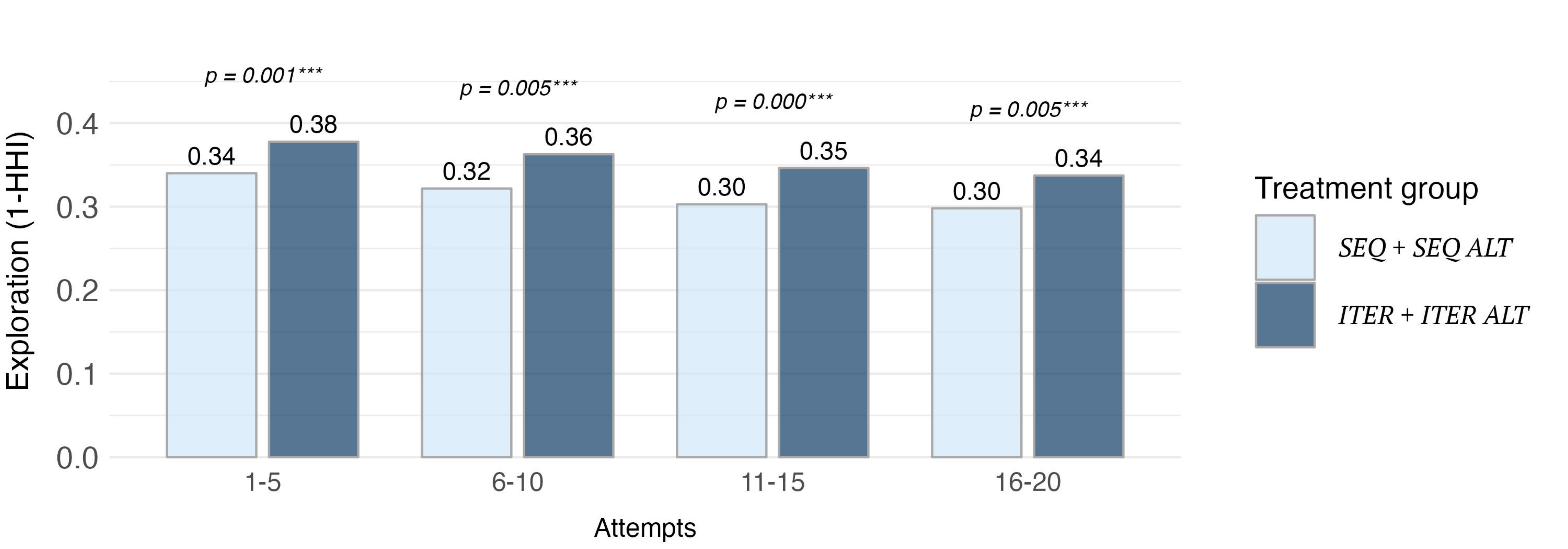}}
{Lemonade Stand Task: Exploration Behavior by Attempt (Product Component)\label{fig:HHL:treat}}
{$p-$values are based on rank sum tests. $^*p<0.10;~ ^{**}p<0.05; ~^{***}p<0.01$.}
\end{figure}

We first examine exploration behavior (1$-$HHI) in each treatment and after each attempt. We focus on the Product component for clarity of exposition; however, similar insights are obtained for the Market component (See Appendix EC.3.4. for more detailed analysis). The treatment differences are shown in Figure \ref{fig:HHL:treat}. As can be seen, exploration declines gradually as participants go from the initial attempts towards the end of the task. This aligns with intuition: early on, subjects perform a broad search of the solution space, while later attempts are used towards a narrower fine-tuning. Additionally, the figure shows that subjects explore significantly more (i.e., examined solutions are less concentrated) in the iterative workflow relative to sequential workflow. These differences are statistically significant (at $p<0.01$ in all four blocks). Thus, an iterative workflow leads to more exploratory behavior throughout the task. 

\begin{figure}[bt!]
\FIGURE
{\includegraphics[width=\textwidth]{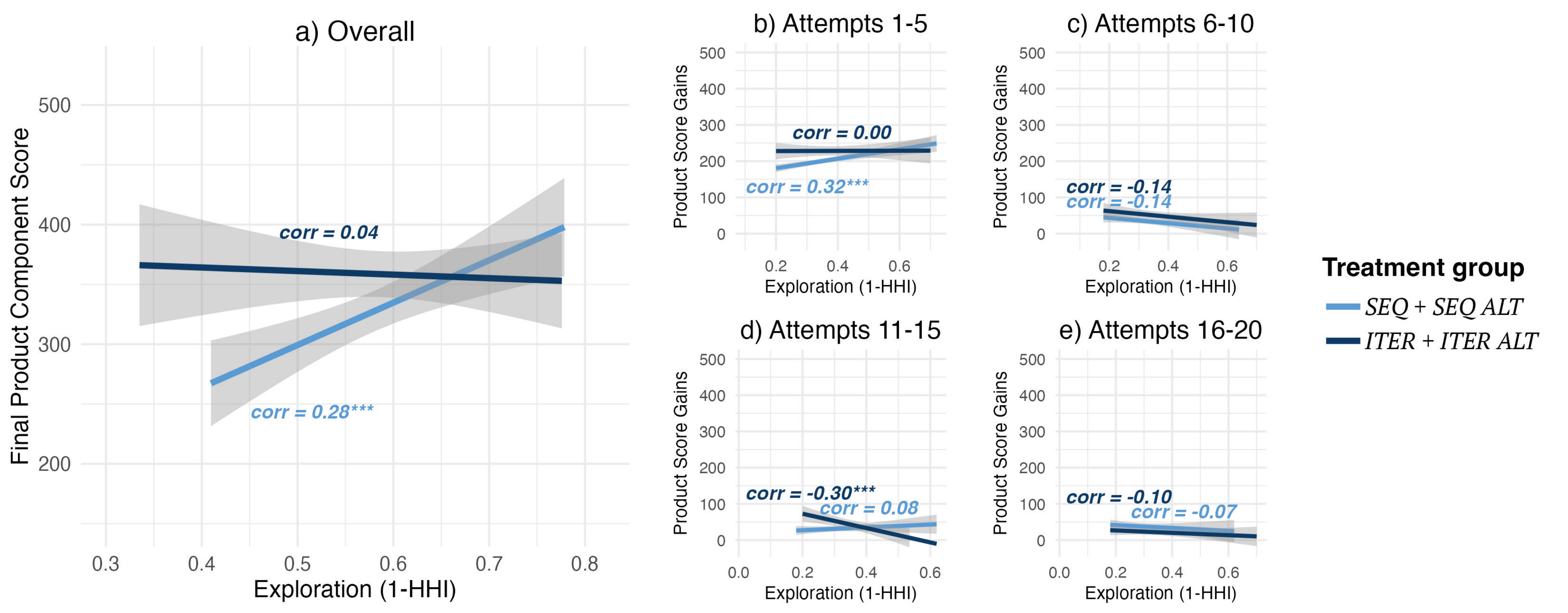}}
{Lemonade Stand Task: Exploration Behavior and Product Component Performance\label{fig:scatter:plots}}
{Figures show linear fit with 95\% CI, as well as Pearson correlation coefficients for each subgroup. $^*p<0.10;~ ^{**}p<0.05; ~^{***}p<0.01$.}
\end{figure}

To further understand the effects of exploration behavior on performance in Figure \ref{fig:scatter:plots}, we examine its relationship with productivity gains. Panel (a) shows the following: while participants in the sequential workflow (\textit{SEQ + SEQ ALT}, light blue) generally achieve lower Product component scores than those in the iterative workflow (\textit{ITER + ITER ALT}, dark blue), there is a strong positive correlation between exploration and performance for sequential participants ($r = 0.28, p < 0.01$). This suggests that when sequential workflow participants engage in high levels of exploration, they can achieve performance levels comparable to their iterative counterparts. This relationship is particularly pronounced early on in the time horizon (panel b, attempts 1-6), where exploration shows a significant positive correlation with score gains for sequential participants ($r = 0.32, p < 0.01$). Interestingly, while sequential workflow may lead to underexploration (i.e., too narrow search), we also find that iterative workflow may lead to too broad exploration. In particular, panel d (attempts 10-15) shows a significant negative correlation between exploration and score gains for the iterative group ($r = -0.30, p < 0.01$). That is, a portion of participants in the iterative workflow continue exploring later on in the time horizon, and that can hurt their performance. While this is not sufficient to offset the early advantage achieved by iterative workflow, it nevertheless suggests that both workflows can result in calibration errors, and that such errors may occur at different times of the work process. 

Together, these findings highlight an important distinction in behaviors between sequential and iterative workflows. The sequential approach leads to a quicker onset of myopic, ``greedy'' hill-climbing search behaviors, which in turn can lead to poor performance. Conversely, iterative workflow facilitates greater exploration and a more dispersed search for the best solution, which helps workers arrive at superior results in less time. This analysis also suggests a practical remedy: encouraging workers in sequential workflows to explore more extensively during early stages may help reduce the performance gap between the two workflow approaches.

\section{Conclusions\label{sec:discussion}}
This section concludes our investigation by presenting an integrated discussion of the results, discussing managerial implications and future directions. 
\subsection*{Integrated Summary}
Table \ref{tab:summary} summarizes our results. Our main finding is that iterative workflow consistently outperforms sequential workflow across multiple innovation environments with distinct structural characteristics. Indeed, iterative workflow achieved higher average performance in three of four experimental tasks. The benefits of transitioning to the iterative approach are both statistically significant and economically meaningful, with performance gains of up to 28 percent. 

Our analysis also reveals several boundary conditions. First, workflow effects were minimal when searching for the best solution on a smooth solution landscape. Second, while iterative workflow achieved a statistically significant advantage during the initial work phase, these benefits were diminished or even reversed in the later work phases. Therefore, when we used either the sum or maximum of component scores as outcome measures, iterative and sequential workflows performed similarly. Finally, iterative workflow failed to outperform sequential workflow when participants began with the easier task component, suggesting that appropriate task sequencing (easy $\rightarrow$ difficult) can serve as a potential remedy for sequential workflow.

While the ability to iterate and switch between tasks helped productivity, we found that this was not explained by improved time allocation. Indeed, in the Scrabble task, performance was the same whether the time spent in each task was exogenously imposed to be equal (\textit{ITER EQUAL FREEZE} treatment) or endogenously determined by the worker (\textit{ITER }and \textit{ITER FREEZE} treatments). This (null) result is surprising given that a less constrained action set should improve productivity. However, this result is consistent with the growing body of work that finds that more autonomy may not always improve performance in complex tasks, such as product design \citep{kagan2018}, project selection and abandonment \citep{long2020}, and time and effort allocation \citep{lieberum2022}. We contribute to this literature, clarifying that while certain types of autonomy, such as the ability to switch between task components, can be helpful, other types, such as flexible time allocation, may not.

\begin{table}[b!]
\TABLE
{Summary of Results\label{tab:summary}}
{\small 
\begin{tabular}{lL{2.7cm}C{2.4cm}C{3.4cm}C{5.6cm}}
    \toprule
      \textbf{Section}    & \textbf{Task} & \textbf{Nature of task} & \textbf{Preferred workflow} & \textbf{Boundary conditions} \\
    \midrule 
    \addlinespace
   \textsection 4,\textsection 7.1 & Scrabble & Idea generation \& recombination & Iterative & \multicolumn{1}{L{5.6cm}}{\multirow{8}{*}{\parbox[t]{5.6cm}{1) Significant productivity differences in first period, not in second; 2) Minimal effects of workflow when performance is evaluated based on the sum or the maximum of component scores. 3) Minimal effects of workflow when participants start with the easier of the two components. }}} \\ \addlinespace
    \textsection 5 & Scrabble with pre-formed words & Recombination  & Iterative &  \\
    \addlinespace
%\hdashline
\addlinespace
   \textsection 6,\textsection 7.1 & Lemonade stand, rugged landscape & Recombination  & Iterative &  \\ \addlinespace
   \textsection 6   & Lemonade stand, smooth landscape & Recombination  & No significant differences &  \\ \addlinespace 
    \bottomrule 
\end{tabular}}
{}%
\end{table}%

\subsection*{Practical Implications}
Our results echo the day-to-day experience of our industry partner, Allianz’ cybersecurity division (see \textsection1). Similar to our experimental setup, the products and services developed within their group are best characterized as ``weakest link'' systems -- a complex software solution performs only as well as its weakest component. Our key findings about learning patterns also match their observations. Since they started using the Agile approach, Allianz' management has seen increased frontloading of work, reduced idleness in the initial stages, and decreased rework in the later stages of development. These real-world observations are broadly consistent with our experimental findings about how iterative workflows change work patterns over time.

The Allianz example also helps clarify when managers should adopt iterative rather than sequential workflow. Sequential workflow falls behind at the start but improves as workers learn, whereas iterative workflow delivers steady gains from the outset. This distinction becomes particularly relevant when the weakest component determines overall performance. For example, when launching a mobile app, a confusing interface or a slow-loading feature can undermine user adoption regardless of how well other elements function. Similarly, in autonomous vehicle development, any subsystem failure makes the entire platform unusable. In these cases, stable performance across all features or components favors the iterative approach. In contrast, when performance evaluation is additive, or when the best-performing feature drives success, a sequential workflow can achieve comparable results. The sequential approach may thus be well-suited for resource-constrained firms, such as early-stage technology startups, where focusing on excelling in one key dimension (e.g., core functionality or user acquisition) may be desirable. A lean startup might deliberately sequence their product development, focusing on the core offering before expanding to related features. Thus, the choice between iterative and sequential workflows ultimately depends on whether success requires consistent performance across all components or breakthrough performance in specific areas.

\subsection*{Outlook}
The experimental tasks used in our study represent only a subset of all innovation settings. First, we have focused on settings in which tasks are independent (i.e., do not require integration) and are only linked through a payoff function. Future work may expand on our experiment by adding a third stage where participants must integrate completed components. Second, our study focused on individual activities with precise and immediate performance feedback.  Interpersonal dynamics, both in collaborative teams found in R\&D, and in competitive contexts like innovation tournaments, might offer interesting variations. Third, our experiments involve relatively short work intervals (several minutes), which leads to minimal fatigue effects. Workflow and task-switching may have different effects on behavior when it occurs over longer periods of time. Fourth, future research (both analytical and empirical) can further explore how the behaviors identified in our experiments influence effort and performance in more competitive settings such as innovation tournaments \citep{terwiesch2008}, where workers allocate their time and attention across multiple simultaneous assignments \citep{korpeouglu2022, kizilyildirim2022}, or in models of entrepreneurial time allocation \citep{yoo2016}.

%\bibliographystyle{informs2014}

%%%%%%%%%%%%%%%%%
\end{document}